\documentclass[aps,prl,10pt,twocolumn,superscriptaddress,amsmath,amssymb,showkeys]{revtex4-2}
\usepackage{graphicx}   %Package for loading pictures from folder
\graphicspath{{Plots/}}
\usepackage{hyperref}   %Creates the links for all references, formulas and pictures
\usepackage{siunitx}
\usepackage{float}
\usepackage{booktabs}
\makeatletter
\let\saved@includegraphics\includegraphics
\AtBeginDocument{\let\includegraphics\saved@includegraphics}
\makeatother

\usepackage{xcolor}
\definecolor{myBlue}{rgb}{0.1,0.1,0.7}

\newcommand{\vint}{v_0}
\newcommand{\vuc}{\bar{v}}
\newcommand{\Deff}{D_\mathrm{eff}}

\graphicspath{./images/}
\usepackage{xcolor}
\usepackage{bm}        
\usepackage{amsfonts}
\usepackage{verbatim}
\usepackage{tikz}

\usepackage{dcolumn} 
\def\d{\mathrm{d}}

\begin{document}
\title{Reaction Delays Boost and Rectify Swimming in  Activity Landscapes}
\title{Time-Delayed Motility  Boosts and Rectifies Swimming in Activity Landscapes}
\title{Time Delays can Boost, Rectify, and Reverse Swimming in Activity Landscapes}
\title{Delayed Motility can Boost, Rectify, and Reverse Swimming in Activity Landscapes}
\title{Reaction Delays  Boost, Rectify, and Reverse Motility in Activity Landscapes}

\author{Constantin Rein}
\email{rein@itp.uni-leipzig.de}
\affiliation{Leipzig University, Faculty of Physics and Earth Sciences, Institute for Theoretical Physics, Brüderstraße 16, 04081 Leipzig}
\author{Klaus Kroy}
\affiliation{Leipzig University, Faculty of Physics and Earth Sciences, Institute for Theoretical Physics, Brüderstraße 16, 04081 Leipzig}
\author{Viktor Holubec}
\email{viktor.holubec@matfyz.cuni.cz}
\affiliation{Department of Macromolecular Physics, Faculty of Mathematics and Physics, Charles University, 18000 Prague, Czech Republic}

\date{\today}
\begin{abstract}
Local detailed balance restricts transport by self-propulsion in static activity landscapes. We show that a delayed speed adaptation (``delayed motility/kinesis'') generically breaks this symmetry, inducing directed transport in asymmetric periodic motility profiles and enhancing diffusion in symmetric ones. Both effects occur for run-and-tumble, active Brownian, and inertial active particles, without requiring potential interactions, walls, imposed gradients, higher dimensions, or translational diffusion.  Their magnitudes, and even current reversals, are conveniently tuned via the delay time, which establishes delayed motility as a versatile generic and experimentally accessible mechanism for autonomous self-steering and transport control in motile-active-matter circuity.
\end{abstract}

\keywords{time-delayed motility, kinesis, microswimmers, reaction delay, activity landscapes}

\maketitle

Motile active matter requires energy sources to fuel its self-propulsion. Uneven fuel distributions therefore naturally render the propulsion speed position-dependent. In biology, such speed modulation without orientational bias is commonly known as (chemo- or photo-)kinesis to distinguish it from directed transport along external gradients, so-called taxis~\cite{eisenbach2004chemotaxis}. Also artificial microswimmers, such as laser-heated or catalytic Janus particles in spatially varying illumination or concentration fields exhibit kinesis~\cite{mijalkov_engineering_2016,Soker2021,holubec2025delayedactiveswimmervelocity,Topfer2025}. The central question we address is whether and how a static spatial activity landscape can, by itself, induce autonomous directed transport, in absence of a global gradient or auxiliary interactions. 

The motion of ordinary overdamped active particles with position-dependent speed $v(\mathbf r)$  and stochastic orientation $\hat{\mathbf n}(t)$ is typically modeled by a Langevin equation
\begin{equation}
\dot{\mathbf r}(t)=v\left(\mathbf r(t)\right)\hat{\mathbf n}(t)
+\sqrt{2D}\,\boldsymbol{\eta}(t),
\label{eq:EQMarkov}
\end{equation}
where $D$ is the translational diffusivity and 
$\boldsymbol{\eta}$ unbiased unit-variance Gaussian white noise. This description both encompasses Markovian Run-and-Tumble Particles (RTPs) 
and Active Brownian Particles (ABPs)~\cite{cates_2013}. For RTPs, $\hat{\mathbf n}(t)$ undergoes Poissonian jumps with rate $\Omega$; for ABPs, rotational Brownian motion with diffusivity $\Omega$. In two dimensions, the unit vector $\hat{\mathbf n} = (\cos \theta, \sin \theta)^\intercal$ boils down to an angle variable $\theta(t)$.

A local detailed-balance condition~\cite{metzger2026exceptionsratchetprincipleactive,SupplementalMaterial} forbids current rectification or ``ratcheting'' of such microswimmers in static activity landscapes unless broken, in at least two dimensions, by translational diffusion ($D>0$)~\cite{rein_force-free_2023}, external forces~\cite{angelani_geometrically_2010,Pietzonka2019}, or  interacting ensembles~\cite{stenhammar_light_induced_2016}. 
In this Letter, we show that delayed motility provides a generic mechanism to break this symmetry, independently of dimensionality, interactions, external forces, or translational noise, and that it can generate efficient (if not perfect) rectification in asymmetric yet periodic activity landscapes. To underscore the generality, we compare discrete reaction delays and exponential memory in overdamped ABPs with overdamped and inertial RTPs. In all cases, a net stationary current can be induced and its magnitude controlled by the delay.  Discrete delays optimize rectification and maximize transport. They moreover allow for current reversals, when commensurate with traversal times across characteristic features of the activity landscape. Besides such tunable directed transport in \emph{asymmetric} periodic activity profiles, delayed motility can enhance diffusion in \emph{symmetric} activity landscapes.

%For minimal but experimentally relevant models, we derive analytical results that elucidate the underlying mechanisms and substantiate our main conclusions, which we further corroborate by Brownian dynamics simulations in more complex settings.

\textit{Delayed-motility models} --- For active particles with noticeable memory or inertia, the time-scale separation between particle motion and velocity relaxation is lost, and Eq.~\eqref{eq:EQMarkov} must be generalized to
\begin{equation}
\dot{\mathbf r}(t) 
= \mathbf v[\mathbf r(t),\hat{\mathbf n}(t)]
+ \boldsymbol{\zeta}(t),
\label{eq:EQNonMarkov}
\end{equation}
with a causal functional $\mathbf v[\mathbf r,\hat{\mathbf n}]$   of the past trajectory of position and orientation, and, in general, a colored noise $\boldsymbol{\zeta}(t)$. We consider the three most common situations.  

ABP, RTP: For overdamped ABP or RTP, the thermal white noise from Eq.~\eqref{eq:EQMarkov} can be retained, $\boldsymbol{\zeta} \to \sqrt{2D}\,\boldsymbol{\eta}$, while memory $\mathbf v[\mathbf r(t),\hat{\mathbf n}(t)] =  v[\mathbf r(t)] \hat{\mathbf n}(t)$ induced by the swimming mechanism is encoded in a speed functional
\begin{equation}
 v[\mathbf r(t)] = \int_{-\infty}^t \!\! \d t'\, \Gamma(t-t')\, v\left({\mathbf r}(t')\right)\,.
\label{eq:vOverdamped}
\end{equation}
Chemotactic bacteria or artificial Brownian microswimmers, which adjust their speed via a hidden (e.g., biochemical) Markov process, exhibit \emph{exponential memory}, $\Gamma(t) \to \tau^{-1}\exp\left(-t/\tau\right)$~\cite{eisenbach2004chemotaxis,Topfer2025}.
\emph{Discrete delays}, for which the memory kernel is a $\delta$-function, $\Gamma(t) \to \delta(t-\tau)$ (corresponding to infinitely many hidden degrees of freedom~\cite{Loos2021}), are more suitable to model feedback control of robots~\cite{mijalkov_engineering_2016} or artificial Brownian microswimmers~\cite{holubec2025delayedactiveswimmervelocity}. Our discussion of both idealized cases should inform expectations for arbitrary interpolations between them. 

\emph{Inertial active particles} --- In some situations, particle and/or solvent inertia cannot be neglected. For the former, one writes
$
\tau \ddot{\mathbf r}(t) + \dot{\mathbf r}(t)
= v[\mathbf r(t)]\,\hat{\mathbf n}(t)
+\sqrt{2D}\,\boldsymbol{\eta}(t),
$
with $\tau$ given by particle mass over friction coefficient~\cite{LowenIABP}. Integration over time yields delayed dynamics of the form~\eqref{eq:EQNonMarkov} (Sec.~S5~\cite{SupplementalMaterial}) with the functionals
\begin{equation}
\mathbf v[\mathbf r(t),\hat{\mathbf n}(t)]
=
\frac{1}{\tau} \int_{-\infty}^t \!\!\d t'\,
e^{-(t-t')/\tau}\, 
v\!\left[\mathbf r(t')\right]
\hat{\mathbf n}(t')
\label{eq:Inertia}
\end{equation}
and $\boldsymbol{\zeta}(t) = \sqrt{2D} /\tau\int_{-\infty}^t \d t'\,
e^{-(t-t')/\tau}\,\boldsymbol{\eta}(t')$ for the swim velocity and noise, where any prior memory $v\!\left[\mathbf r(t)\right]$ and the stochastic force $\boldsymbol\eta$ have been bestowed with inertia by convolution with $\exp(-t/\tau)$. In contrast to the overdamped active models with memory in Eq.~\eqref{eq:vOverdamped}, the swim direction $\hat{\mathbf n}$ enters the memory integral in Eq.~\eqref{eq:Inertia}, and the stochastic force $\boldsymbol{\zeta}(t)$ exhibits exponential temporal correlations on top of possible prior correlations of $\boldsymbol\eta$.
In the following, we are content with inertial RTPs without prior memory, i.e., $v\!\left[\mathbf r(t)\right] = v\!\left(\mathbf r(t)\right)$ and white noise $\boldsymbol\eta$.

%Inertial active particles: In some situations, particle and/or solvent inertia cannot be neglected. For the former, one writes
%$
%\tau \ddot{\mathbf r}(t) + \dot{\mathbf r}(t)
%= v[\mathbf r(t)]\,\hat{\mathbf n}(t)
%+\sqrt{2D}\,\boldsymbol{\eta}(t),
%$
%with $\tau$ given by particle mass over friction coefficient~\cite{LowenIABP}. Integration over time yields delayed dynamics of the form~\eqref{eq:EQNonMarkov} (Sec.~S5~\cite{SupplementalMaterial}) with the functionals 
%\begin{equation}
%\mathbf v[\mathbf r(t),\hat{\mathbf n}(t)]
%=
%\int_{-\infty}^t \!\!dt'\,
%\Gamma(t-t')\, 
%v\!\left[\mathbf r(t')\right]
%\hat{\mathbf n}(t')
%\label{eq:Inertia}
%\end{equation}
%and $\boldsymbol{\zeta}(t) = \sqrt{2D} %/\tau\int_{-\infty}^t dt'\,
%e^{-(t-t')/\tau}\,\boldsymbol{\eta}(t')$ for the swim velocity and noise, where any prior memory $v\!\left[\mathbf r(t)\right]$ and the stochastic force $\boldsymbol\eta$ have been bestowed with inertia by convolution with $\exp(-t/\tau)$. In contrast to the overdamped active models with memory in Eq.~\eqref{eq:vOverdamped}, the swim direction $\hat{\mathbf n}$ enters the memory integral in Eq.~\eqref{eq:Inertia}, and the stochastic force $\boldsymbol{\zeta}(t)$ exhibits exponential temporal correlations on top of possible prior correlations of $\boldsymbol\eta$.
%In the following, we are content with inertial RTPs without prior memory, i.e., $\Gamma(t) = \delta(t)$ and white  $\boldsymbol\eta$.

%%%%%%%%%%%%%% fig 1 %%%%%%%%%%%%%%%%%%%%
\begin{figure}
\centering
\includegraphics[width=\linewidth]{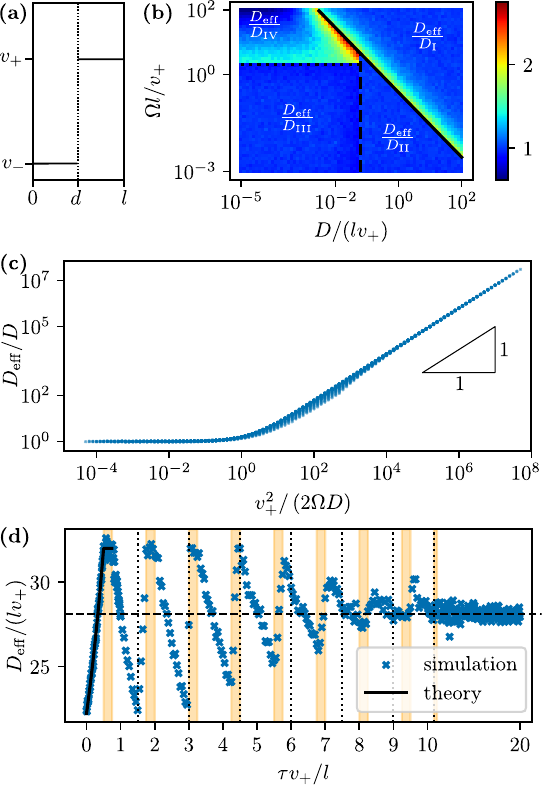}
\caption{Memory-enhanced diffusion of RTPs with discrete delay $\tau$ in a symmetric one-dimensional periodic activity landscape from Brownian dynamics simulations.
\textbf{(a)} Unit cell of the periodic 2-step activity profile with high- and low-activity intervals of widths $d$ and $(l - d)$ with nominal speeds $v_-$ and $v_+ = v_-/2$, respectively. Effective diffusion coefficient $\Deff$ for the optimum ``leapfrog'' delay $\tau=d/v_+$ as function of \textbf{(b)} the reduced tumbling rate $\Omega l/v_+$ and thermal diffusivity $D/(l v_+)$ (relative deviations from the predictions of Tab.~\ref{tab:1} are color-coded and their parameter regimes delimited by lines), \textbf{(c)} active Péclet number $v_+^2/(2\Omega D)$, and \textbf{(d)} reduced delay time $\tau v_+/l$ for $D = 0$ and $\Omega l/v_+ = 10^{-2}$ in regime III, with an analytical prediction from Eq.~\eqref{eq:Deff} (solid line). Dotted lines and shaded regions mark limit cycles with minimum and maximum net unit-cell traversal speed \textcolor{black}{$\vuc$}, respectively. Simulations were started with maximum initial speed $v(t \le 0) = v_+$, iterated up to times $t=10^5 l/v_+$ ($10^8$ steps) and $5\cdot10^6 l/v_+$ ($5\cdot10^9$ steps), and averaged over $3\times10^3$ and $1.5\times10^4$ independent realizations, for \textbf{(b--c)} and  \textbf{(d)}, respectively.}
%Raw data for panels \textbf{(b)} and \textbf{(c)} are available in Fig.~S1~\cite{SupplementalMaterial}.}
\label{fig:1}
\end{figure}
%%%%%%%%%%%%%%%%%%%%%%%%%%%%%%%%%%%%%%%%%

%%%%%%%%%%%%%% tab 1 %%%%%%%%%%%%%%%%%%%%
\begin{table}
\begin{tabular}{cccc}
    \toprule
    \mbox{}  & Constraints & Main Mechanism &  $\vuc$  \\ \midrule
    I & $D>\frac{v_+^2}{2\Omega}$ & thermal diffusion & negligible \\ \addlinespace
    II & $v_-^2 \tau < D < \frac{v_+^2}{2\Omega} $ &  $D$-blurred $v(x)$  & $\frac{v_- d}{l} +v_+ (1-\frac{d}{l})$ \\ \addlinespace
    III & $\Omega < \frac{v_+}{l-d}$, \textcolor{black}{$D<v_-^2\tau$} & $\tau$-delayed $v(x)$ &  Eq.~(S2.1) \\ \addlinespace
    IV & $\Omega > \frac{v_+}{l-d}$, $D<\frac{v_+^2}{2\Omega}$ & $\Omega$-blurred $v(x)$ & $\frac{v_- v_+\sqrt{d^2 + (l-d)^2}}{\sqrt{d^2 v_+^2 + (l-d)^2 v_-^2}}$ \\ 
    \bottomrule
\end{tabular}
\iffalse
 \begin{tabular}{|c|c|c|c|}
    \mbox{}  & Constraints & main mechanism &  $\vuc$  \\ \hline
    I & $D>\frac{v_+^2}{2\Omega}$ & thermal diffusion & negligible \\
    II & $v_-^2 \tau < D < \frac{v_+^2}{2\Omega} $ &  $D$-blurred $v(x)$  & $\frac{v_- d}{l} +v_+ (1-\frac{d}{l})$ \\
    III & $\Omega < \frac{v_+}{l-d}$, $D<\frac{v_-^2}{\tau}$ & $\tau$-delayed $v(x)$ &  (S2.1) \\
    IV \hfill & \hfill $\Omega > \frac{v_+}{l-d}$, $D<\frac{v_+^2}{2\Omega}$ \hfill & \hfill $\Omega$-blurred $v(x)  \hfill $ \hfill & $\frac{v_- v_+\sqrt{d^2 + (l-d)^2}}{\sqrt{d^2 v_+^2 + (l-d)^2 v_-^2}}$ \hfill %\frac{\sqrt{l^2+(d-1)^2}v_+ v_-}{\sqrt{l^2v_-^2+v_+^2 (d-1)^2}}$ \hfill
 \end{tabular}
 \fi
 \caption{Predictions for the active diffusivity and its dominant mechanism for various parameter regimes defined by constraints on the tumbling rate $\Omega$, bare translational diffusivity $D$, and delay time $\tau$. Substitution of $\vuc$ in Eq.~\eqref{eq:Deff} yields the effective diffusion coefficients $D_\text{I}$–$D_\text{IV}$ for Fig.~\ref{fig:1}(b).}
 \label{tab:1}
\end{table}
%%%%%%%%%%%%%%%%%%%%%%%%%%%%%%%%%%%%%%%%%

\textit{Enhanced diffusion} --- In the absence of additional forces, no systematic directed transport occurs,  \emph{in spatially symmetric activity profiles}~\cite{Soker2021,holubec2025delayedactiveswimmervelocity,Topfer2025}. Nevertheless, as we now demonstrate, generalizing recently reported  results for ABPs with exponential memory~\cite{garces2026universaltransportactivecolloids}, velocity memory can profoundly modify the swimmer's effective diffusivity.
To elucidate the mechanism, consider the one-dimensional periodic activity landscape in $\mathbf r \to x$ of Fig.~\ref{fig:1}(a), with a unit cell in which the propulsion speed is a step function switching between  $v(x)=v_+$ and $v_- < v_+$, in intervals of width $l-d$ and $d$, respectively.
Let us first concentrate  on the degenerate limit $v_-=0$ and $D = 0$, where the low-activity regions are absorbing ($\dot{x}=0$) and permanently trap Markovian ($\tau=0$) RTPs and ABPs, so that their long-time diffusivity
$\lim_{t\to\infty}\,\langle x^2(t)\rangle/{2t}$ vanishes.
With a discrete reaction delay, however, strongly persistent  RTPs with a very low tumbling rate $\Omega \ll v_+/d$, can leapfrog over a trap within the time $d/v_+<\tau$ and keep going for time $\tau - d/v_+$ until taking a pause of duration $d/v_+$.  Provided they have landed in the active region, they can then resume their motion. 
Notably, even in the joint limit $d,\tau\to0$, any nonzero absorbing interval traps Markovian swimmers whereas delayed swimmers may pass. 
%The complementary degenerate limit of an activity step %profile with $v(\mathbf r)=0$ everywhere except for %infinitesimally narrow spikes of speed $v_+$, further %underscores the mechanism. Indeed, RTPs initialized on %any such  motility spike in a periodic lattice with %the exact lattice constant $v_+\tau$, are %ballistically transferred to the next motility spike, %where they pause for a delay time $\tau$ before their %next leap. 
Similarly, in the absence of orientational and translational noise, a fine-tuned reaction delay alone can yield sustained (though unstable) transport across a one-dimensional periodic activity landscape, even when the activity vanishes almost everywhere (provided that one starts at an active site). Delay may thus qualitatively change the fluctuations.

We now turn to a more generic setting, in which perfectly absorbing states are eliminated either by a finite translational diffusivity $D>0$ or by a non-vanishing minimum activity ($0<v_-<v_+$). As expected, the advantage of a time-delay, namely to facilitate leaps across the low-speed regions that dominate the net traversal speed $\vuc$ across the unit cell, tends to be washed out by noise. Yet, delayed speed adaptation in a symmetric periodic motility profile may still manifest itself in a considerably \emph{enhanced effective long-time diffusivity} 
\begin{equation}
\Deff(\tau)=D+\frac{\vuc^2}{2\Omega}, %\frac{\bar{v}^2(\tau)}{2 \Omega} \,.
\label{eq:Deff}
\end{equation}
that augments the thermal diffusivity $D$ (the only contribution for negligible activity) by a self-propulsion contribution, characterized by a constant effective active speed $\vuc$~\cite{cates_2013}. Table~\ref{tab:1} lists the dominant diffusion mechanisms with their parameter regimes and predictions for $\vuc$.  For sufficiently weak thermal diffusion ($D < v_-^2 \tau/2$) that only marginally blurs the underlying motility pattern $v(x)$, a  ``leapfrog-and-tumble mode'' can considerably boost the effective diffusivity (regime III). For large persistence ($\Omega \ll v_+/l$), when several unit cells are traversed between consecutive reorientation events, $\Deff$ is found to grow linearly in $v_+^2/(2\Omega D)$, Fig.~\ref{fig:1}(c). 
%this follows form the the one-dimensionality of the data in the fig1c
%In the corresponding parameter regime III in Fig.~\ref{fig:1}(b), the unsteady propulsion effectively amounts to swimming at a constant average propulsion speed  $\bar{v}(\tau)$, resulting in the form
% \begin{equation}
%D_A =D_{\mathrm{III}} \equiv \frac{\vuc^2(\tau)}{2\Omega}.
%\label{eq:DA-III}
% \end{equation}
%For all other parameter choices in Tab.~\ref{tab:1}, time-delay effects tend to be washed out by noise.

According to our analytical estimate  in Sec.~S2.A.1~\cite{SupplementalMaterial}, \textcolor{black}{the effective active speed $\vuc$ in Eq.~\eqref{eq:Deff}}
%the net transfer speed $\vuc(\tau)$ in the unit cell 
peaks for $d/v_+<\tau < (l - d v_-/v_+)/v_+$. This result is corroborated by the Brownian-dynamics simulations of RTPs, shown in Fig.~\ref{fig:1}(d). The decaying oscillations and eventual saturation of $\Deff(\tau)$ can be understood as follows.
For a delay time $\tau =d/v_+$ and $D=0$, the active motion quickly enters a stable limit cycle~\cite{strogatz2000nonlinear}, where the perceived motility $v(t)$ is time-periodic with period \textcolor{black}{$T_{\mathrm{min}} = l/\max{\vuc}$} and thus in resonance with the spatially periodic motility pattern $v(x)$, resulting in a \emph{maximally boosted random motion}. Trajectories obtained without delayed reaction $(\tau =0)$ are trivially also on a limit cycle, but with period 
\textcolor{black}{$T_{\mathrm{max}} = l/\min{\vuc}$,}
which corresponds to the slowest possible \textcolor{black}{mean speed $\vuc$}. %mean speed $\vuc$. 
For longer delay times  $\tau = n T_\mathrm{max}$ and $\tau=\tau_m + n T_\mathrm{min}$, with $n \in \mathbb{N}$ and $\tau_m \in [d/v_+ ,(l - d v_-/v_+)/v_+]$, stable limit cycles that maximize $\vuc$ and $\Deff$ can only be attained from fine-tuned initial conditions. Since every tumbling event effectively restarts the process of convergence with random initial conditions, and also longer limit cycles with mean speeds anywhere between the extreme values may exist, these resonances and the corresponding oscillations of $\Deff$ between its largest and smallest possible value get increasingly washed out for growing $\tau$. %, at large $\tau$, where $\Deff$ saturates at a value corresponding to an averaged motility landscape.

\textit{Rectified transport in one dimension} ---  To obtain directed transport, an \emph{asymmetric} periodic activity profile is required. A minimal model extends the 2-step unit cell from above by a third interval of width $d$ with a motility  $v(x)=\vint$, intermediate between $v_+$ and $v_-$ (Fig.~\ref{fig:2}a).
Consider again first the noiseless ($D$, $\Omega \to 0$) ``trapping limit''  ($v_-=0$)  for the ``leapfrog'' delay time $\tau=d/v_+$, in which the three-step activity staircase acts as a rectifying diode:  the swimmer can still only just traverse the trapping interval when ascending but gets trapped when descending the staircase. The stationary current is obtained by averaging over the ascending and descending unit-cell traversal speeds $v_\uparrow$, $v_\downarrow$, for which explicit expressions in terms of $v_\pm$, $v_0$, $\tau$ and $l$, $d$ can be found in Sec.~S2.B~\cite{SupplementalMaterial}. %(Adding the translational and rotational noise back in would merely deteriorate the synchrony of the leapfrog motion across the absorbing intervals and thereby reduce the directed transport.) 
It reaches an upper bound 
\begin{equation}
\langle \dot{x} \rangle = \lim_{t\to \infty} \frac{\langle x(t) - x(0) \rangle}t
= \frac{v_\uparrow-v_\downarrow}{2}
= \frac{v_+/2}{1+\tau v_+/l}
\end{equation}
for $\vint\lesssim v_+$ (maximum forward current) and $v_-\to 0$ (no retrograde current). Which in turn approaches the optimum $v_+/2$ achievable by any sorting ratchet, for $\tau \to 0^+$. Since the current vanishes for $\tau=0$ and $\vint= v_+$, this upper bound is singular and never truly achievable in a real physical setup, much like the Carnot efficiency for a thermodynamic cycle. Also notice that, with the delay time fixed to the exact leapfrog value $\tau= d/v_+$, the limit $\tau\to 0$ is obtained for $d\to 0$, corresponding to a rectifying zero-width activity hole that only just suppresses the motion in one direction, entirely. 

\begin{figure}
\includegraphics[width=\linewidth]{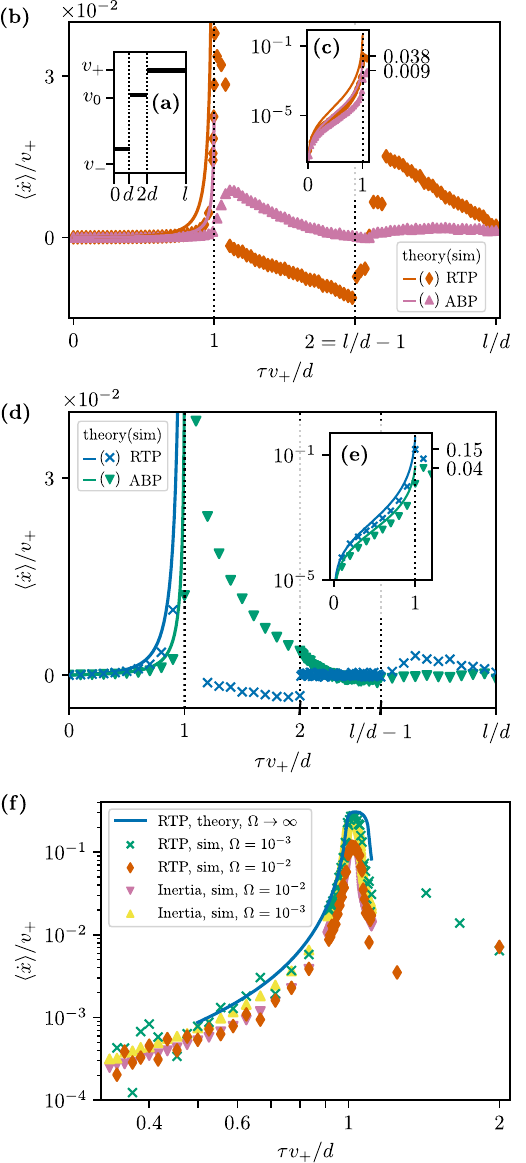}
\end{figure}
\begin{figure}
\caption{Memory-induced transport in asymmetric activity landscapes. 
\textbf{(a)} Unit cell of the 3-step activity profile with propulsion speeds $v_-$, $\vint$, and $v_+$ in sections of length $d$, $d$, and $(l-2d)$ of the unit cell, respectively.
\textbf{(b-f)} Net current $\langle\dot{x}\rangle$ versus delay time $\tau$ for $\vint = 0.9v_+$, $v_-=10^{-3} v_+$ and $D=0$, in \textbf{(b)} and \textbf{(d)} for RTPs and ABPs with $\Omega=10^{-2}l/v_+$ and with $d=l/3$ and $d=l/10$, respectively. Insets \textbf{(c)}, \textbf{(e)} magnify the interval $\tau v_+ = [0,d]$, where the current attains its maximum. Analytical estimates (solid lines) for $\tau\le d/v_+$ and $\Omega \ll v_+/l$ closely mimic our Brownian dynamic simulations (symbols). %In (a), the current for RTPs  reversesat $\tau = d/v_+$ and $2d/v_+$. 
In \textbf{(d)}, the current behaves qualitatively as in \textbf{(b)}, except for an additional interval $2d/v_+<\tau<(l-d)/v_+$, where it gradually decays to zero for ABPs and vanishes for RTPs. 
\textbf{(f)} RTPs with exponential delay and inertial RTPs [Eq.~\eqref{eq:Inertia}] (for several values of $\Omega$) coincide for $\Omega \to 0$. Analytical approximations from  Sec.~S2.B~\cite{SupplementalMaterial} (lines) take only one-step memory into account.  Simulation times were $10^9 \, l/v_+$ ($10^{12}$ time steps) (b-e, RTP), $10^8\, l/v_+$ (b-e, ABP) and $4\cdot10^6\, l/v_+$ (f, RTP). Inertial motion was obtained by piecewise solution (in each velocity interval), over a total time of $10^9 \, l/v_+$.}
\label{fig:2}
\end{figure}
%%%%%%%%%%%%%%%%%%%%%%%%%%%%%%%%%%%%%%%%%
For finite $d$, $v_-$, and $v_+ - \vint$, the current remains substantial for the optimum delay time $\tau = d/v_+$ just long enough for the swimmer to leapfrog over the least active region, namely $\max\langle\dot{x}\rangle$ up to about $20\%$ of $v_+$, according to  Fig.~\ref{fig:2}
%(a-d), 
(cf.~Fig.~S3~\cite{SupplementalMaterial}),
while current reversals occur for larger $\tau$.
%More precisely, the current is negative for $(2-\vint/v_+)d/v_+ < \tau < 2d/v_+$, \textcolor{red}{when descending swimmers can leap across the slowest regions (speed $v_-$, $v_{\mathrm{0}}$) and thereby rapidly return to the fast region with speed $v_+$, whereas an ascending swimmer spends most of its time in the intermediate-speed region (speed \( v_{\mathrm{0}}\))
More precisely, the current turns negative for $(2-\vint/v_+)d/v_+ < \tau < 2d/v_+$ provided that $d<l/3$ and $v_-$ is  sufficiently small for the swimmer to spend at least the delay time $\tau$ in the most active region (Sec.~S2.B.1~\cite{SupplementalMaterial}). Descending swimmers can then traverse the intermediate-activity and the least active regions at speeds $v_+$ and $v_0$, respectively. In contrast, ascending swimmers spend most time in the intermediate-activity region.
Such \emph{``retrograde rectification''}  is generally weaker than the forward rectification for \(0<\tau<(2-\vint/v_+)d/v_+\), yielding peak currents of a few percent of \(v_+\). For \(2d/v_+<\tau<(l-d)/v_+\), the leap distance exceeds \(2d\) and both ascending and descending swimmers spend the same time in the \(v_+\)-region, yielding no rectification and a vanishing net current, although possibly substantially enhanced diffusion. For \((l-d)/v_+<\tau<l/v_+\),  ascending swimmers effectively bypass the \(v_-\)-region, which traps descending swimmers for extended periods, yielding again a positive current. Finally, for \(\tau>l/v_+\), existence and direction of rectification depend sensitively on the values of \(v_-\), \(\vint\), and \(v_+\), precluding analytical predictions, see Fig.~S3~\cite{SupplementalMaterial}.
%%%%%%%%%%%%%%%

In general, rectification is reduced when resonances between the travel distance during a characteristic delay time and geometric features of the activity profile are weakened, either by noisy locomotion or by mechanisms blurring the geometry of the activity landscape. For an instructive paradigmatic example of a continuous activity landscape that yields similar (though lower) currents as the step profile, see Sec.~S3~\cite{SupplementalMaterial}.

Added translational and rotational noise ($D$, $\Omega>0$) not only disrupts the leapfrog synchronization across passive intervals, as observed for \(D_{\rm eff}\) in Fig.~\ref{fig:1}(c), but also introduces additional relevant length scales, such as the persistence length \(v_+/\Omega\). Their interplay with the geometry (\(d\), \(l\)) of the activity profile determines whether rectification persists, through mechanisms similar to those controlling \(\Deff\) and exemplified in Fig.~\ref{fig:1}(d). Details can be found in Sec.~S2.B.2, Tab.S1 and Figs.~S4--S6~\cite{SupplementalMaterial}.
Similarly, the ABPs' orientation-dependent velocity as well as added memory and inertia in case of RTPs all impair synchronization. However, aside from a reduced current magnitude and missing current reversals in the inertial and exponential-memory cases, the qualitative behavior remains similar to that discussed above for bare RTPs. In Figs.~\ref{fig:2}(a,d), current reversals are barely discernible for ABPs, but see Fig.~S7, Fig.~\ref{fig:2}(f) for inertial and exponential-memory RTPs, and Secs.~S2--S4~\cite{SupplementalMaterial}.

\textit{Rectification in two dimensions} --- In contrast to the one-dimensional case considered here, a two-dimensional ratchet yields directed transport even for vanishing delay ($\tau=0$)~\cite{rein_force-free_2023}. Introducing a finite delay gives rise to current reversals whenever the characteristic distance traveled over the delay time $\tau$ matches one of the characteristic length scales of the activity profile; see Figs.~S10--S11~\cite{SupplementalMaterial}. %Current reversals are not observed for ABPs in one-dimensional activity landscapes, but they become possible in higher dimensions, as we demonstrate in Sec.~S5~\cite{SupplementalMaterial}, using a two-dimensional ratchet that can sustain a nonzero current, even for Markov dynamics ($\tau=0$) if \(D>0\)~\cite{rein_force-free_2023}. 
%As in one dimension, current reversals occur when the characteristic distance traveled per delay time $\tau$ matches one of the length scales in the activity profile; see Figs.~S9–S10~\cite{SupplementalMaterial}.

%To substantiate this claim, we investigated the role of a discrete time delay in the two-dimensional force-free active ratchet of Ref.~\cite{rein_force-free_2023}. 
%\textcolor{red}{In contrast to the one-dimensional system considered here, the two-dimensional ratchet supports directed transport even for vanishing delay time~$\tau$~\cite{rein_force-free_2023}.}Generalizing the model to include a nonzero delay enables control over both the magnitude and the direction of the resulting current [Sec.~S5 in Ref.~\cite{SupplementalMaterial}].

%Similar memory-induced current reversals are expected to be a generic feature of delay–driven transport. 

\textit{Discussion} --- We have established delay-induced current rectification and enhanced diffusion as generic features of simple paradigmatic active-matter models with spatially inhomogeneous activity. Net currents are possible, even in one-dimensional static activity landscapes, where directed motion is strictly forbidden in the absence of memory~\cite{rein_force-free_2023,metzger2026exceptionsratchetprincipleactive}, demonstrating that a retarded response alone can lift fundamental symmetry constraints on transport in active systems. We have explained how both rectification and enhanced diffusion originate from resonances between the transport distance during the delay time and length scales intrinsic to the activity profile. These resonances are progressively washed out by increased noise, more broadly distributed memory, or orientation-dependent transport of the swimmers. Importantly, for the canonical active Brownian particle model (ABP), which describes most experimental realizations, both discrete and exponentially distributed delays generate effects of sufficient magnitude to be observable under standard thermal noise conditions, although a reduction of fluctuations, for instance via confinement to quasi two-dimensional geometries, should facilitate the experiments. %Integrating out the orientational degrees of freedom in standard swimmer models or the viscoelastic response of a complex solvent produces effective Langevin equations for the position variable with memory and or colored (multiplicative) noise, thereby generically breaking the fluctuation-dissipation relation. Consequently, such dynamics may also be expected to produce directed currents or enhanced diffusion. 
Overall, the robustness of the delayed response mechanisms we described underscores the generic character of their consequences in motile active matter. In practice, our results imply that (chemo-/photo-)kinesis, while incapable of generating directed transport in its idealized instantaneous form, should generically produce systematic motion in asymmetric periodic activity landscapes, once finite response times are accounted for. Delayed kinesis due to feedback, relaxation, signal processing or transmission, and inertia~\cite{Loos2021,holubec2025delayedactiveswimmervelocity}, and the ensuing currents in heterogeneous environments, should thus be generic features for a wide variety of biological and synthetic motile agents. Which, in turn, suggests that microorganisms might deliberately employ delay-navigation. Also in many experimental settings, delays are not only inevitable but tunable. Our results thus establish a general paradigm of autonomous-transport control via temporal programming and a versatile toolkit for a future engineering of ``soft active circuitry''.

\textit{Data Availability} --- No publicly available research data or software support this manuscript. The data can be reconstructed from the equations in Ref.~\cite{SupplementalMaterial} and are available from the authors upon reasonable request.

%\textit{Author Contributions} --- Conceptualization, Methodology: C.R., V.H., K.K.; Formal analysis: C.R., V.H.; Investigation, Software, Data curation, Visualization: C.R.; Supervision: V.H.; Writing --- original draft: C.R.; Writing --- review \& editing: C.R., V.H., K.K.

\bibliography{references}

\end{document}